\documentstyle[12pt,aasms]{article}
\newcommand{\pard}{\partial}

\newcommand{\etal}{et~al.}
\newcommand{\Ai}{\mbox{Ai}}
\def \paczy{Paczy{\'n}ski}
\begin{document}

%

\title{Femtolensing: Beyond the Semi-Classical Approximation}

\author{Andrew Ulmer\altaffilmark{1} and
	Jeremy Goodman\altaffilmark{2}}
\affil{Princeton University Observatory,
       Peyton Hall,
       Princeton, NJ~08544,
       USA}
\altaffiltext{1}{E-mail: andrew@astro.princeton.edu}
\altaffiltext{2}{E-mail: jeremy@astro.princeton.edu}

\begin{abstract}
Femtolensing is a gravitational lensing effect in which the magnification
is a function not only of the positions and sizes of the source and lens, but
also of the wavelength of light. Femtolensing is the only known effect of
$(10^{-13}-10^{-16} M_{\sun})$ dark-matter objects and may possibly be
detectable in cosmological gamma-ray burst spectra. We present a new and
efficient algorithm for femtolensing calculations in general potentials. The
physical-optics results presented here differ at low frequencies from the
semi-classical approximation, in which the flux is attributed to a finite
number of mutually coherent images. At higher frequencies, our results agree
well with the semi-classical predictions. Applying our method to a point-mass
lens with external shear, we find complex events that have structure at both
large and small spectral resolution. In this way, we show that femtolensing
may be observable for lenses up to $10^{-11}$ solar masses, much larger than
previously believed. Additionally, we discuss the possibility of a search for
femtolensing of white dwarfs in the LMC at optical wavelengths.

\end{abstract}


\section{Introduction}

The possibility of interference effects in gravitational lensing
has been considered by several authors (\cite{Man:81}; \cite{ss:85};
\cite{dw:86}; \cite{pf:91}; \cite{g:92}; \cite{spg:93}).
Of particular interest are diffractive variations in flux with frequency
$\nu$ when the source, lens, and observer occupy fixed positions.
If the difference in time delay between a pair of images satisfies
$\nu\Delta t\gg\gg 1$, the fringe spacing is
$\Delta\nu\approx \Delta t^{-1}\ll\nu$.
We may call $\nu\Delta t\gg 1$ the ``semi-classical'' regime, because
diffractive phenomena are produced by mutually
coherent images whose positions, magnifications, and time delays
can be determined using geometric optics.
On the other hand, if $\nu\Delta t\lesssim 1$,
regions of the lens plane other than
the geometric-optics images contribute importantly to the flux.
The semi-classical approach then breaks down, and one must use the methods
of physical optics.

If observed, interference effects would reveal dark-matter objects
in a mass range to which few other tests are sensitive.
The characteristic time delay produced by a lens of mass $M$ is
\begin{equation}
\label{char_time}
\Delta t(M)=2GM/c^3= R_{\rm Sch}/c,
\end{equation}
where $R_{\rm Sch}$ is the Schwarzschild radius.
Hence the condition $\nu\Delta t\sim 1$ is equivalent to
$\lambda\sim R_{\rm Sch}$.
Since the Schwarzschild radius of the Sun is $\approx 3 \mbox{km}$,
broad-band fringes ($\Delta\nu\sim\nu$) require decidedly sub-stellar
but nevertheless macroscopic lensing objects.
In particular, Gould (1992) has shown that lens masses
$M\sim 10^{-16}-10^{-13} M_{\sun}\sim 10^{17}-10^{20}\mbox{g}$
could produce observable fringes
in gamma-ray burst spectra at energies $E\sim 1\mbox{MeV}$ ($\lambda\sim
10^{-10}\mbox{cm}$).
Because the angular separation of the images produced by such a
lens is $\sim 10^{-15}$ arc sec, Gould has coined the name
``femtolensing'' for this phenomenon.
On the other hand, the probability that a randomly placed and
cosmologically distant point-like source should be lensed is
$\sim\Omega_{\rm lens}$, where the latter is the mean mass density
in lensing objects expressed as a fraction of the critical density
$3H_0^2/8\pi G$ (\cite{pg:73}).
Given that $\sim 10^3$ gamma-ray bursts have been detected to date
(CGRO Science Report 157 1994),
a single well-established case of femtolensing would indicate that
objects $\sim 10^{-16}M_{\sun}$ contribute significantly to the mass
density of the universe.

Even if copious lenses exist in the appropriate mass range, visible
fringes can be seen only if the (incoherent) source is smaller than the
Fresnel length $(\lambda D)^{1/2}$, where $D$ is the distance.
For source redshifts of order unity, this translates to
$R_{\rm source}\le 10^{14}\lambda_{\mbox{cm}}^{1/2}\mbox{cm}$.
An additional constraint requires that in order to have
significant magnification, the source size must be
smaller than the Einstein ring radius. For cosmological distances,
$R_{\rm source}\le 5 \times 10^{8} M/(10^{-16} M_{\sun})$~cm.

The rapid time variability of gamma-ray bursts (GRB)
of 0.2 msec (\cite{bh:92}) as well as the cosmological distances
of $\sim 0.5$~Gpc for bright BATSE bursts found from Log~$N$-Log~$P$
studies (\cite{fe1:93})
suggest, however, that {\it GRB are sufficiently compact}.
There has been some confusion as to whether the appropriate
linear source size should be taken from
$\gamma c\Delta t$ or $\gamma c t$ where $\gamma$ is the bulk Lorentz
factor, $c$ is the speed of light, $\Delta t$ is the smallest time
variation detected, and $t$ is the total event duration.
For many proposed cosmological scenarios, the appropriate measure
is $\gamma c\Delta t$, because the last-scattering-surface
remains at approximately the same radius even though
the relativistic ejecta may reach quite large distances in the course of
the burst.
For $\gamma$'s of 100--300 (e.g. \cite{fe:93}),
the (linear) size of a GRB last-scattering-surface is of order
$5\times 10^{8}$~cm.
Other models predict
emission from patches
on a relativistically
expanding shell which can become extremely large ($\gamma c t$).
In the latter case, it would be difficult to observe femtolensing
because the interference patterns would differ from patch to patch.
An observation of femtolensing could distinguish between the two
scenarios.

Stanek, Paczyn\'nski, \& Goodman (1993, henceforth SPG)
have discussed the possibility that line
features in burst spectra may have been produced
by femtolensing.
Such lines have been seen or inferred in
GINGA data (\cite{Mu:88}, \cite{fe:88}) and
KONUS data (\cite{Maz:81})  and have been
attributed to cyclotron absorption.
Like cyclotron lines, interference fringes would be evenly spaced in
photon energy.
No convincing evidence for lines has yet been seen
in the largest homogeneous data set available, the BATSE experiment
on the Compton Gamma-Ray Observatory,
although its capability of line detection is lower (\cite{T:93}).

The femtolensing calculations cited above
have considered only the simplest possible case, which is
an isolated point-mass lens.
The simplicity and symmetry of such a lens allow the physical
optics problem to be solved in terms of known functions
(\cite{dw:86}).
In the present paper, we present an efficient
physical-optics method for computing frequency-space fringes
produced by general lenses.
We assume that the lensing mass distribution is
confined to a layer thin compared to the observer-source distance
(single-screen approximation).
We also neglect time dependence of the lensing geometry, which is
permissible if the time-delay difference between any pair
of images changes by less than $\nu^{-1}$ during an observation.
The latter assumption is probably justified for femtolensing of
gamma-ray bursts (\cite{g:92}).

For definiteness, and because it is the simplest lens not yet
treated in physical optics, we apply our methods to a single point
mass with external shear.
The computational approach taken, however, would apply equally well to an
arbitrary surface density of lensing mass.
Computational savings are achieved mainly by taking advantage
of the achromaticity of gravitational lensing: that is, the
time delays and excess optical path lengths are independent of
frequency.

The plan of our paper is as follows.
The physical-optics problem is posed in \S II.
We show how the scalar diffraction amplitude at the observer can
be determined as a function of frequency by first calculating
its Fourier transform, which is a function of time delay.
An efficient numerical procedure for finding the latter function is
developed in terms of contour integrals on the lens plane.
The role of the geometric-optics images and the correspondence
with the semi-classical approximation is explained.
In \S III we describe certain details of our numerical implementation of
the method and show that in the case of an
isolated-point-mass lens, our results are
consistent with those already obtained by \cite{dw:86} and SPG
for the isolated-point-mass lens.
In \S IV we present results for the more complex case of a point mass
with external shear, which can produce up to four images.
Finally, \S V briefly summarizes the main points of our approach and prospects
for observing femtolensing.

\section{Formalism}

The quantity of interest for femtolensing is the observed magnification as
a function of frequency of the lensed source relative to the intensity
in absence of a lens.
In scalar diffraction theory,
for the case of a thin screen, this function can be written as
complex square of the amplitude
\begin{equation}
\label{psi}
\Psi(\omega)= C_{\omega} \int^{\infty}_{-\infty}
\int^{\infty}_{-\infty} {\rm d}x {\rm d}y \exp(i\omega\tau),
\end{equation}
where $\omega$ is the photon frequency, $C_{\omega}$ is a normalization
to unit flux in absence of a lens
that varies slowly with $\omega$, $x$ and $y$ are
coordinates in the lens plane, and $\tau$ is the time delay function
(e.g., \cite{bn:86}):
\begin{equation}
\label{gentau}
\tau(\vec r)=
\frac{1+z_l}{c}\left[\frac{d_{os}}{2d_{ol}d_{ls}}(\vec r -\vec r_s)^2
- \frac{2G}{c^2}\int dx'dy'\Sigma(\vec r')\ln|\vec r -\vec r'|\right].
\end{equation}
In the latter formula, $z_l$ is the redshift of the lens; $d_{os}$,
$d_{ol}$, and $d_{ls}$ are angular diameter distances;
$\vec r_s=(x_s,y_s)$ is the point where a direct line from observer
to source would meet the lens plane in the absence of the lens;
and $\Sigma(\vec r)$ is the mass per unit area in the lens plane.

Although our methods are general, we illustrate them by application
to a point mass with sub-critical shear.
If the observer and lens are on the z-axis and the source is slightly
off axis, the time delay function for this case is, in normalized units
(see Appendix A),
\begin{equation}
\label{timedel}
\tau(x,y,\mu,\phi,\theta)=(1+\mu)x^2 + (1-\mu)y^2 - 2\theta[x\cos(\phi) +
y\sin(\phi)] - \ln(x^2+y^2),
\end{equation}
where $\mu$ is the shear, $\phi$ is the angle between shear direction and
that of the displacement of the source from the axis, and $\theta$ is the
angle between the
source and z-axis as measured by the observer in units
of the angle subtended by the Einstein ring.

\begin{figure}
  \plotone{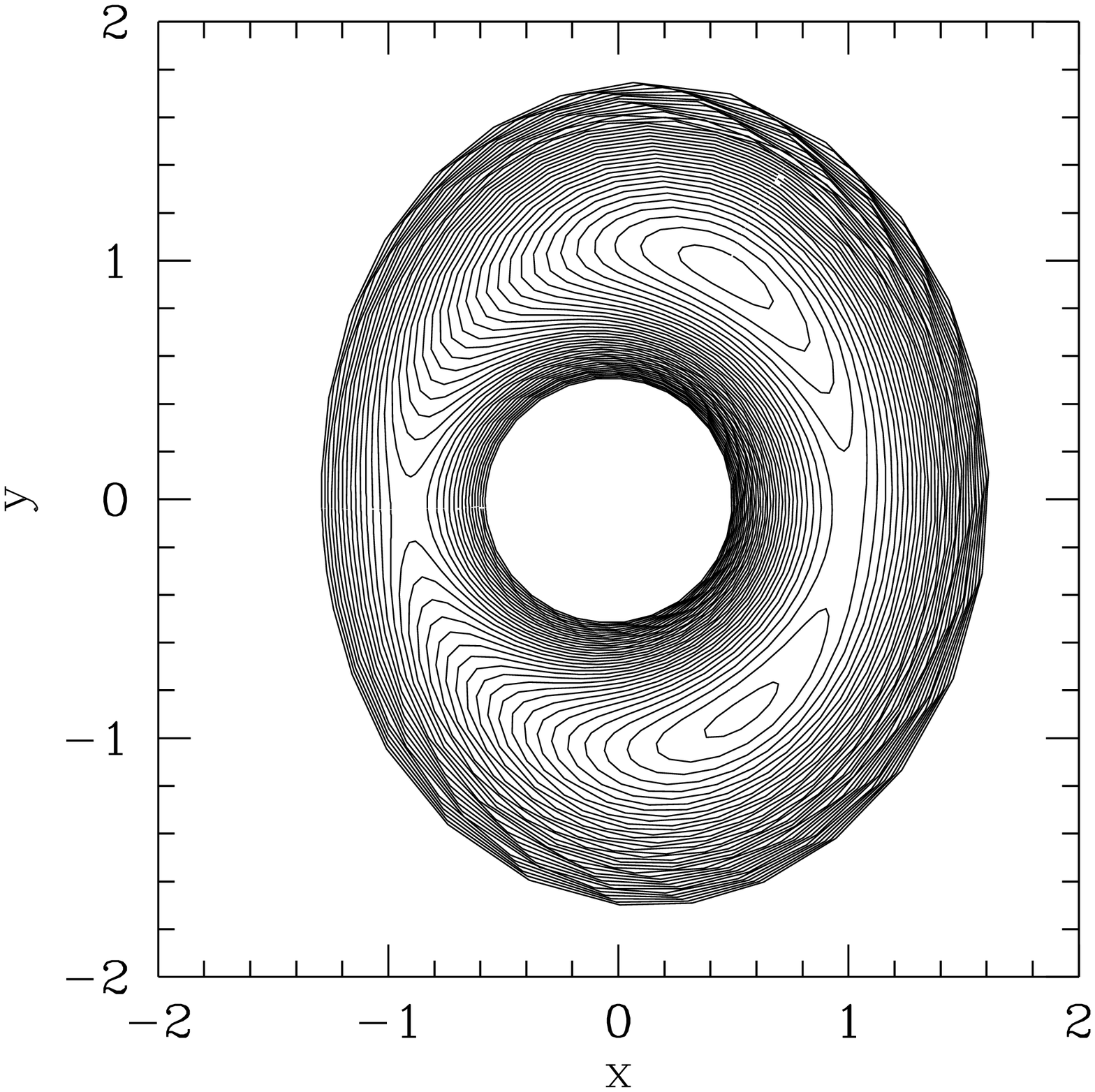}
\caption{This figure shows the contours of
equal time delay in the image plane that
are used to calculate $\tilde{\Psi}(\tau)$ (cf. fig.
\protect\ref{psimap}).
The empty circle in the middle
contains the lens and therefore a logarithmic spike in time delay.
Images form at critical points on
this surface.
Two minima form contours at low time delay on the top and bottom
right of the depicted region.
The contours
merge at a saddle point on the right.
This contour later splits to form
inner and outer circular contours.}
\protect\label{conmap}
\end{figure}

Direct calculation of $\Psi(\omega)$ is a three dimensional problem
in $x, y,$ and $\omega$.
We demonstrate here a method to reduce the problem to
two dimensions with use of Fourier transforms and contour integration.
Dividing Eq.~\ref{psi} by $C_{\omega}$ and taking the Fourier
transform, we define
\begin{equation}
\label{tpsi}
\tilde{\Psi}(t)\equiv \frac{1}{2\pi}\int^{\infty}_{-\infty}
{\rm d}\omega \exp(-i\omega t)
\frac{\Psi({\omega})}{C_{\omega}},
\end{equation}
which can be viewed as a virtual pulse shape in time.
After substitution from Equation \ref{psi}, this equation reduces to
\begin{equation}
\label{tpsi2}
\tilde{\Psi}(t)= \int^{\infty}_{-\infty} \int^{\infty}_{-\infty}
{\rm d}x~ {\rm d}y~ \delta[\tau(x,y)-t].
\end{equation}
Therefore, the contribution to $\tilde{\Psi}(t)$ comes from curves
of constant time delay; an example of such contours is shown in Figure
\ref{conmap}.

Eq. \ref{tpsi2} can be evaluated as a contour integral.
Since
$\tilde{\Psi}(t)\rm{d}t$ is the area between the curves defined by
$ \tau(x,y) = t$
and $ \tau(x,y) = t+\rm{d}t$, and the distance between them is
$\rm{d}t/\|\vec{\nabla}\tau\|$,
\begin{equation}
\label{line1}
\tilde{\Psi}(t) = \oint \frac{{\rm d}s}{\|\vec{\nabla} \tau(x(s),y(s))\|},
\end{equation}
where $\rm{d}s$ is the arc length along the curve defined below.

There will in general be more than one of these contours, though it is
straightforward to find them all as is described in the next
section.
The sum of the integrals (\ref{line1}) over all contours $\{\gamma_k\}$
can be expressed as
\begin{equation}
\label{line2}
\tilde{\Psi}(t)= \sum_k | \oint_{\gamma_k} du|
\end{equation}
if the arc parameter $u$
is defined through the differential equations
\begin{equation}
\label{u}
{\pard x\over \pard u} = -{\pard \tau\over
\pard y}~~~\mbox{and}~~~{\pard y\over \pard u} = {\pard \tau\over \pard x},
\end{equation}
because
\begin{equation}
\label{u2}
{\pard s\over \pard u} = \pm\sqrt{{\left( \pard x\over \pard u \right)^2} +
{\left( \pard y\over \pard u \right)^2}} =  \pm{||\vec{\nabla}\tau||}.
\end{equation}
Notice that equations (\ref{u})
are nonsingular near critical points if $\tau(\vec r)$ is smooth.

\begin{figure}
  \plotone{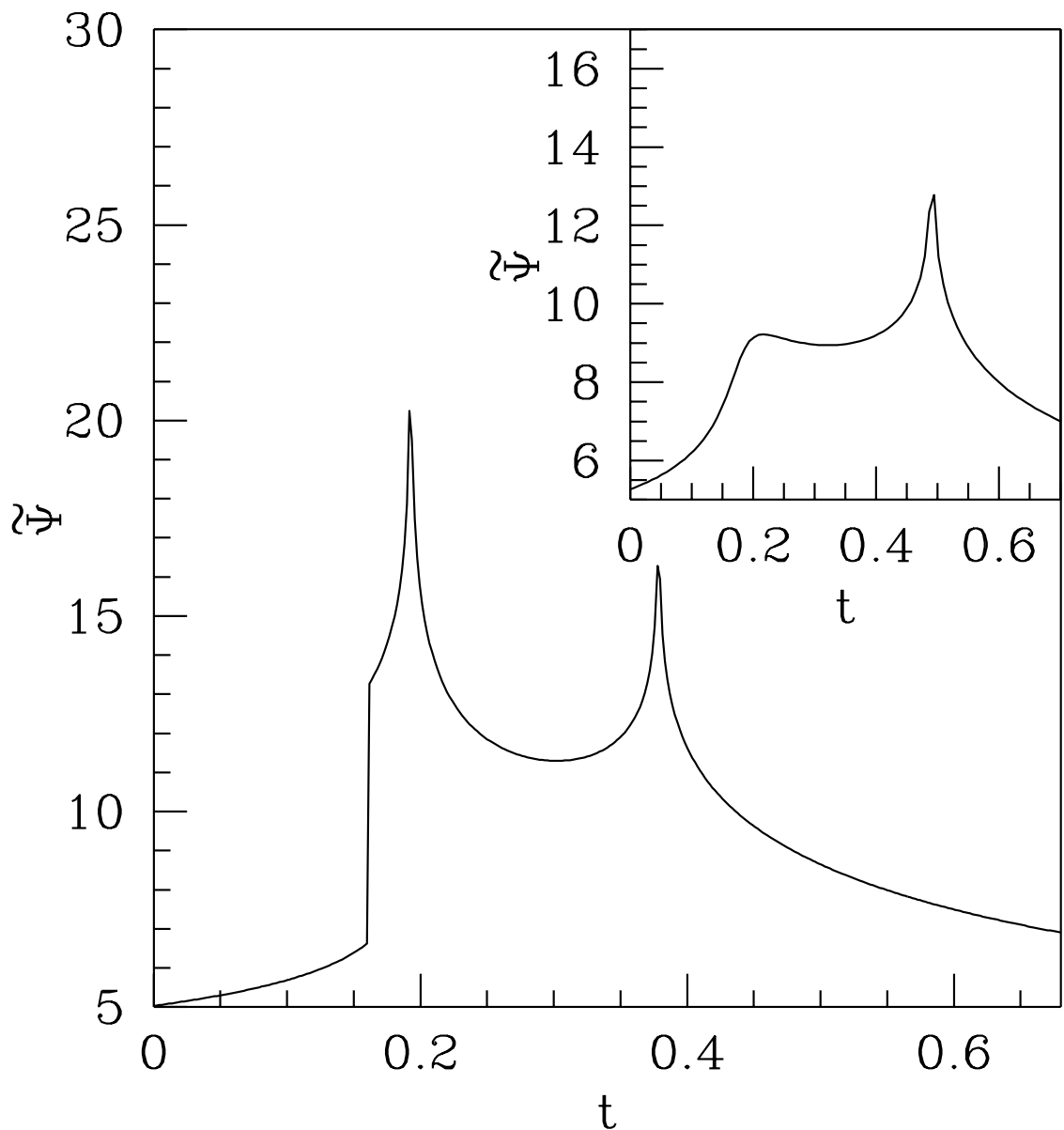}
\caption{Figure \protect\ref{psimap}
shows $\tilde{\Psi}(t)$, which yields
the magnification amplitude, $\Psi(\omega)$ when Fourier transformed.
In this four image geometry $\tilde{\Psi}(t)$ has four
singularities (though the first occurs at minimum time
delay).
Two minima at low time delay cause discontinuities.
Two saddle points at higher time delay form logarithmic spikes.
The inset shows $\tilde{\Psi}(t)$ in a case with only two images
as the source approaches a caustic.
The function resembles that of the four image case.
This transition region is
unique to the wave nature of the physical optics calculation and contrasts with
the semi-classical approximation which has abrupt qualitative
changes between two and four images.}
\protect\label{psimap}
\end{figure}

We find $\tilde{\Psi}(t)$ is a smooth function except where $t$ is
equal to a critical value of the time delay function so that
$\|\vec{\nabla}\tau\|=0$ (an example of $\tilde{\Psi}(t)$ is shown in
Figure \ref{psimap}).
It is at these critical valued time delays that
images form.
These singularities are calculated in Appendix B.
For minima or maxima, there is a discontinuity in
$\tilde{\Psi}(t)$ at the critical time, $t_i$, of
\begin{equation}
\label{minmax}
\lim_{\epsilon \rightarrow 0} ( \tilde{\Psi}(t_i+\epsilon) -
\tilde{\Psi}(t_i-\epsilon) = \pm 2\pi \left[{\pard^2\tau\over
\pard x^2}{\pard^2\tau\over \pard y^2} -
\left({\pard^2\tau\over \pard x\pard y}\right)^2\right]^{-{1\over 2}}.
\end{equation}
In the case of a saddle point, $\tilde{\Psi}(t)$ diverges logarithmically near
$t_i$, so that
\begin{equation}
\label{saddle}
\tilde{\Psi}(t) = -2 \ln |t-t_i|\left[
\left({\pard^2\tau\over \pard x\pard y}\right)^2 -
{\pard^2\tau\over \pard x^2}{\pard^2\over
\pard y^2}\right]^{-{1\over 2}} + \mbox{non-singular part}
\end{equation}
Using a procedure described in the next section, the singularities
can be dealt with and the inverse Fourier transform can be calculated, yielding
the amplitude
\begin{equation}
\label{psi2}
\Psi(\omega)= C_{\omega} \int^{\infty}_{-\infty}
\exp(i\omega\tau)\tilde{\Psi}(t).
\end{equation}

\section{Numerical Implementation}

Except in the simplest circumstances, the method of \S 2 must be applied
numerically.
For a point mass with shear, our algorithm
computes the femtolensed energy spectra in ten to twenty seconds on a Sparc 10.
Computer code which implements the algorithm is available upon request.
For future applications,
we note that the algorithm grows quadratically with the number of lensing
masses.
The computation may be considered in three parts: location of the images,
line integration on constant time delay contours, and calculation of the
inverse Fourier transform of
$\tilde{\Psi}(t)$ to yield $\Psi(\omega)$.

In the case of a point mass lens with shear, the location of the
images can be found analytically, so there is essentially no computational
burden.
However, one could envision more complex applications such as lensing in
a dense field with many lenses
and images where it would be necessary to utilize some image finding scheme.
An efficient image finding algorithm has already been developed
for microlensing calculations (\cite{w:93}).
Note that the femtolensing algorithm increases approximately
linearly with the number of images, so these complicated scenarios may be
tractable.
We find that the
line integration around contours defined in Eqs. \ref{line1}--\ref{u2}
is the most computationally demanding part of the program.

The method by which the contours are found warrants a brief discussion.
An example of these contours are shown, for example, in Figure \ref{conmap}.
The algorithm starts with a contour very close to the minimum time
delay critical point. From there, it moves to larger time delay contours by
taking small steps perpendicular to the contour. This method works as
long as the number or topology of the contours does not change, but this
happens only at the images.
In other words,
the contours are altered only when moving from below the time delay of
an image to above it. When the image is a minimum in the time-delay
surface, a new contour is formed around the minimum. When the contour
is a saddle point, a contour either splits into two, or two contours
merge. One can distinguish between these cases by monitoring the line integral
at contours near each time delay of an image. Maxima are not
encountered in the case of a point mass with shear.

Using the line integrals we determine $\tilde{\Psi}(t)$ for a range of
values of $\tau$, and the computations for one geometry are shown in
Figure \ref{psimap}.
We find generally that one needs approximately 1000
points taken uniformly between the minimum time delay and the point at
which $\tilde{\Psi}(t)$ reaches about one percent of its asymptotic value.
It is impossible to reconstruct the entire detail of the
$\tilde{\Psi}(t)$ curve with such measurements, because for instance, near
the time delay of a saddle point image there is a logarithmic divergence
as discussed in the previous section and Appendix B.
We account for this loss of detail near the critical
time delays by subtracting the analytic forms of the singularities.
(Eqs. \ref{minmax}, \ref{saddle}) and dealing with them separately:
\begin{equation}
\label{sumlines}
\tilde{\Psi}^\prime(t)= \sum_k | \oint_{\gamma_k} du| - \mbox{singular part}
\end{equation}
We then perform a discrete Fourier
transform (DFT) of $\tilde{\Psi}^\prime(t)$, while paying close
attention to edge effects.
Finally, we calculate the exact Fourier transforms of the divergences and
discontinuities and add them to the DFT to obtain $\Psi(\omega)$.

As a test of this procedure, the calculation is carried out in the simplified
case of a point mass without shear and is shown in Figure
\ref{energy1} (top panel).
This calculation agrees with an earlier calculation of the same result
which utilized independent methods (SPG) to the resolution
of the figures presented in this paper.

\section{Results}

\begin{figure}
  \plotone{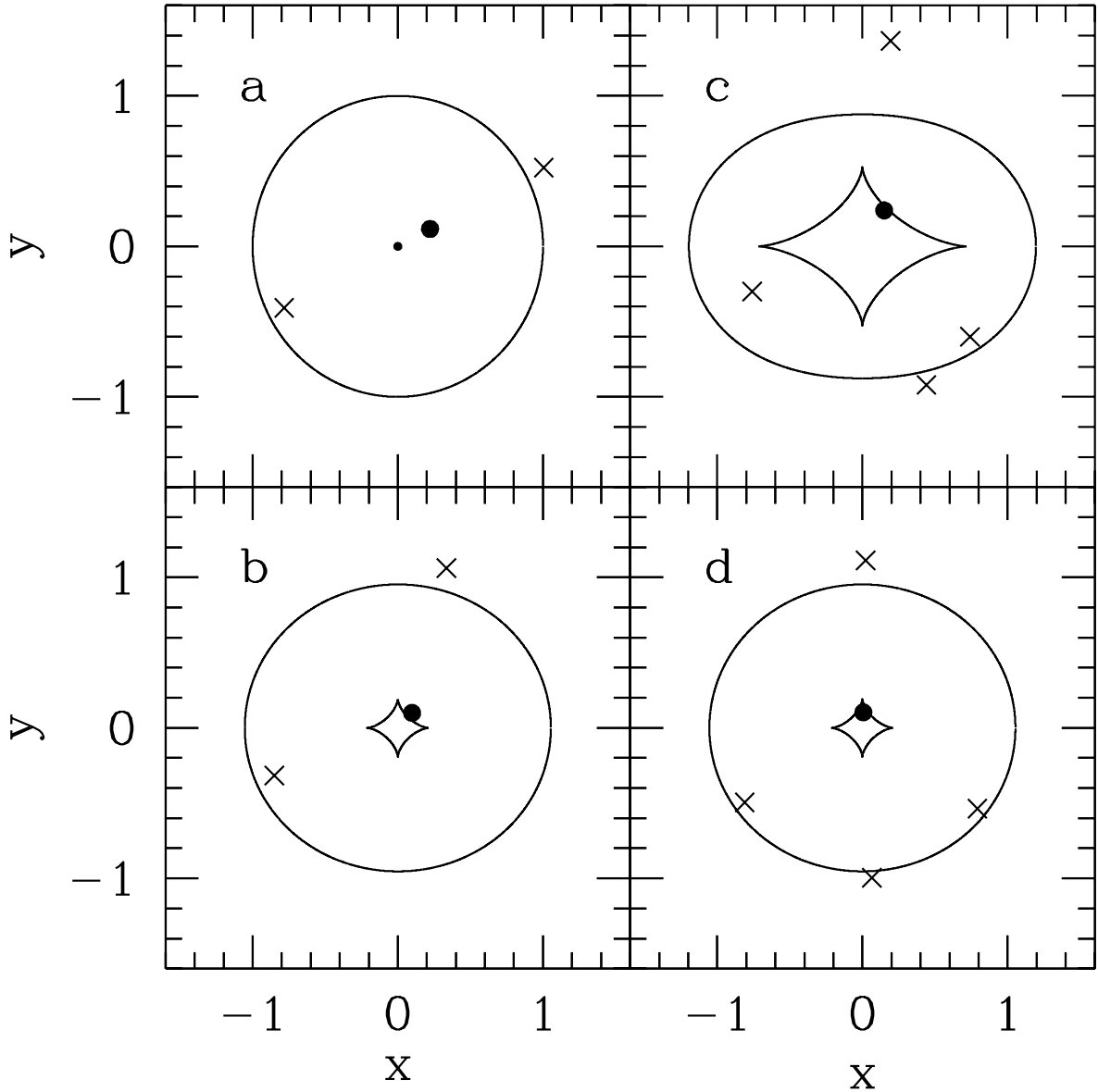}
\caption{Shown in the image plane are the image positions (crosses),
critical curves (circle and ovoids),
projected source positions (large dots),
projected caustics (small dot and diamonds) in dimensionless units for
a variety of geometries. Fig. \protect\ref{caustics}a is a point mass without
shear, so the caustic is the Einstein ring.  Fig. \protect\ref{caustics}b,c,d
have shears of 0.1, 0.3, and 0.1.}
\protect\label{caustics}
\end{figure}

\begin{figure}
  \plotone{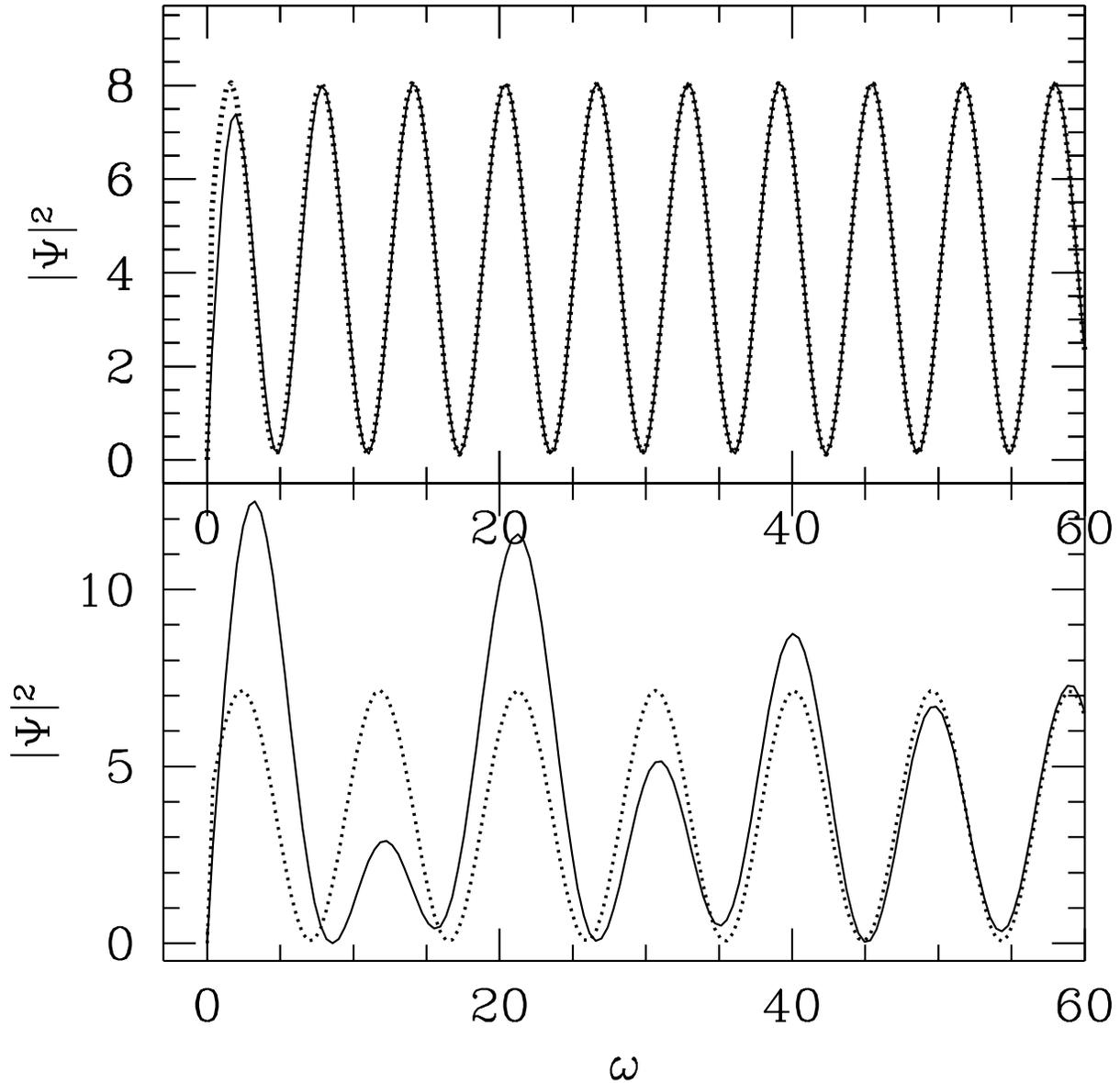}
\caption{Figure \protect\ref{energy1} shows $\|\Psi\|^2$,
the magnification as a function
of dimensionless frequency for the geometries in Fig.
\protect\ref{caustics}a,b
(top and bottom respectively). The solid line is the physical optics
calculation calculated with our algorithm.
The dashed line is the semi-classical, or two-ray,
approximation discussed in appendix B.
When the source approaches a caustic, the energy spectrum cannot be
accurately calculated with the semi-classical approximation
at lower frequencies.}
\protect\label{energy1}
\end{figure}

\begin{figure}
  \plotone{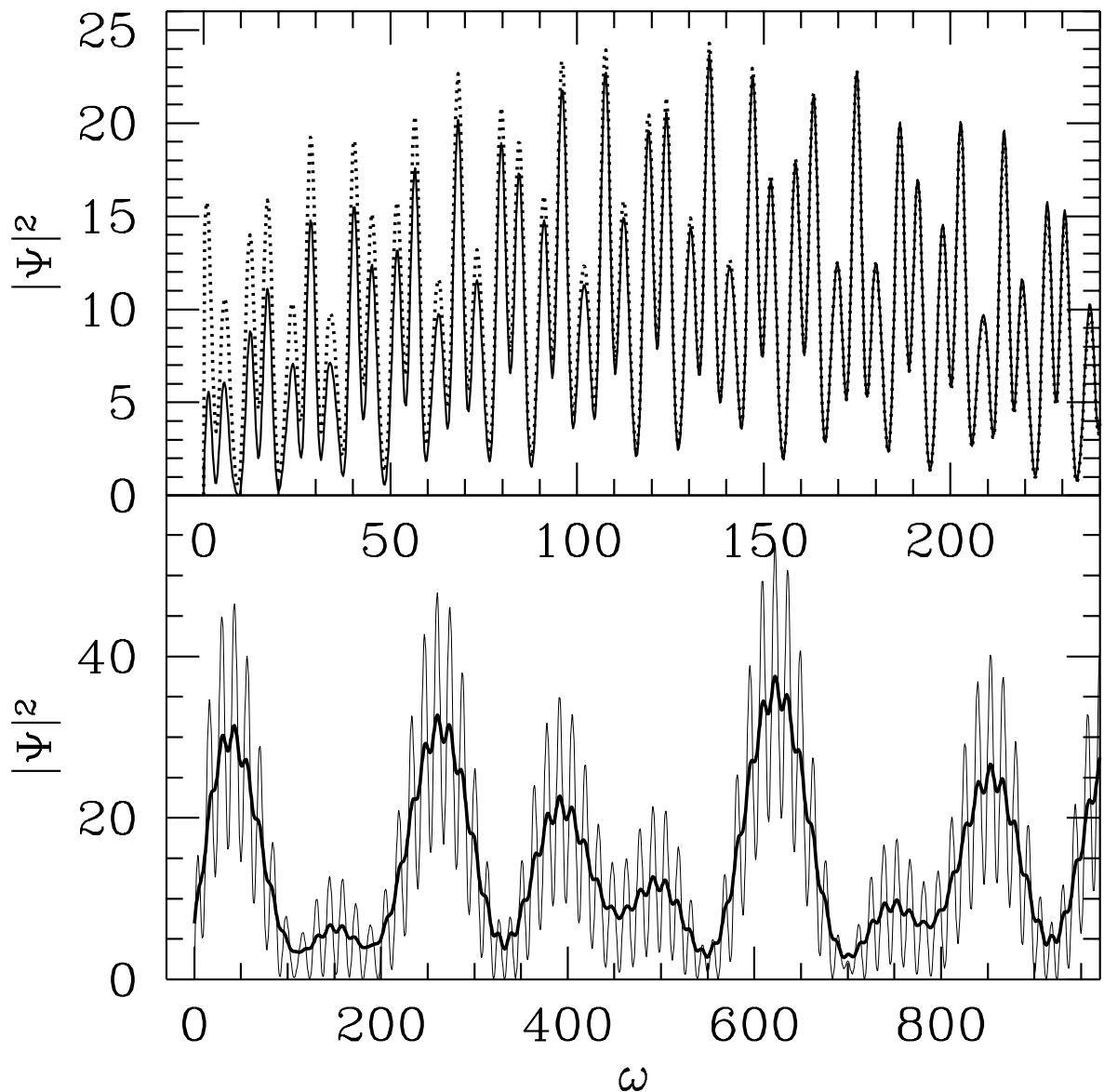}
\caption{This figure shows magnification
as a function of dimensionless frequency
for the geometries in Fig. \protect\ref{caustics}c,d
(top and bottom, respectively).
Complex beat frequencies are created between the time delays of the four
images so that the semi-classical approximation (dashed line in top panel)
is inaccurate. The beat frequencies create features on a much broader
energy range as exemplified by the bottom panel in which a smoothed
magnification curve is superposed on the physical optics calculation.}
\protect\label{energy2}
\end{figure}

Using the new algorithm, we investigated the
interference produced by a variety of lensing geometries for a point
source and lens with shear
(extended sources could be calculated by repeatedly applying the
algorithm for source points chosen, for instance, by Gaussian quadrature
weighings).
Figures \ref{energy1} and \ref{energy2} show the results of calculations for
various source/lens configurations depicted in Figure \ref{caustics}.
The frequency, $\omega$, is given in dimensionless units:
\begin{equation}
\label{omega0}
\mbox{frequency:}~~~\omega \equiv {\tilde{\omega}(1+z_L)R_s
\over c} = \frac{2\pi(1+z_{\rm L})R_s}{\lambda}
\end{equation}
where $R_s$ is the Schwarzschild radius and $\lambda$ is the wavelength.
For example, unity corresponds to photon energy of 1 keV and
a mass of $\sim 5\times 10^{-14} M_{\sun}$ at $z_{\rm L}=0.5$.
A sinusoidal pattern is produced which has a frequency inversely
proportional to the difference in time delay for the two images.
The top panel of Figure \ref{energy1} shows the interference
pattern for a point mass with no shear for
which the semi-classical approximation is an extremely good approximation.

Using our algorithm, external shear can be introduced into the calculation.
We find that when the source approaches the caustic, the physical
optics calculation differs from the semi-classical
approximation. For example, the bottom panel of Figure \ref{energy1} shows
a two image interference pattern for which the first few
interference nodes have largely different amplitudes.
At high frequencies, the semi-classical approximation approaches
the physical optics solution.

The interference patterns become much more complex
when the source lies inside the center caustic region. There,
four images form, and the interference patterns become chaotic.
Figure \ref{energy2} shows the magnification as a function
of frequency for two such cases.
The main characteristics to note are that
the interference patterns can become quite complex with oscillations
with both long and short periods (the bottom panel shows both the physical
optics
magnification and a smoothed magnification).
In short, this means that femtolensing
can be observed over a wider range of frequency space, and therefore,
a wider range of masses than originally believed. Figure \ref{energy2}
shows that the typical interference patterns, corresponding to inverse
time delays, can be hundreds of times larger than the characteristic time
delay (see Eq. \ref{char_time}), so the lens masses can be hundreds
of times larger than those discussed by Gould and SPG.
This short period interference
discussed in previous papers can be seen in the top panel of
\ref{energy1}.
The long period interference occurs when two of the four images
have very close time delays, which is a common result
with four images.
Appendix C shows that in these circumstances
the separation of the long-period fringes increases as
the cube of the total magnification, and the source-size
limit is somewhat more severe than the Fresnel length.

The complex nature of the magnification in Fig. 5 can be understood to some
extent in the semi-classical regime, because there one finds
that the complex amplitude of magnification, $\Psi(\omega)$,
is composed of a set of sinusoidal interference fringes
from each pair of images. For four images,
there are six pairs which each produce sinusoidal interference.
The resulting interference pattern which is the squared sum of these
sine waves will in general be very complex.
The physical optics calculation is qualitatively similar, but
yields different magnifications and different specific structures
at low frequencies.
According to the result (\ref{fringes}) in Appendix C, the first
few long-period fringes are least sensitive to the angular size of
the source.

\section{Discussion}

We find that even in the simplified case of a point mass with external shear,
very complex interference patterns can be formed. In reality, however,
the patterns should be even more complex. In addition to
external shear, one should account for local shear from neighbors for
femtolensing matter in galactic halos.
Complex, many image geometries will result, and as found in the case of
a point mass with external shear, images with smaller-than-characteristic
difference in time delay
(see Eq. \ref{char_time})
will cause surprisingly
long period interference patterns (e.g. Figure \ref{energy2})
which allow for detection of larger masses.
For an external shear of 0.1, we expect these complex interference patterns
to occur 3\% of the time that a source is found inside the Einstein radius.
In general, for a halo  mass
distribution similar to a singular isothermal sphere,
the shear would have a scale length of 5 kpc, so that between 10 and 50 kpc,
the shear would be 0.5-0.1, so these complex patterns would occur much more
often.
Furthermore, there would be a magnification bias (see Appendix C)
towards these complex events, which could likely increase the observed
fraction by an order of magnitude.

Cosmological gamma-ray bursts are considered the best candidates
to show femto-lensing because of their extremely small angular size.
If the line of sight to
a gamma-ray burst passed through a galaxy, and if the dark halo mass
are composed of $\sim 10^{-11}-10^{-16} M_{\sun}$ (we increase the range of
masses by a factor of 100 as a result of the long period events discussed
above) then one could expect
to see, in addition to macro or microlensing, femtolensing effects
as well. The femtolensing could easily mimic
other emission or absorption line processes.
We suggest, then, that if a
spectral absorption or emission feature is definitively observed in a
gamma-ray burst due to femtolensing, that it is likely that the
burst would be macrolensed as well. Additionally, macrolensing
would produce multiple images (bursts) which would likely be
femtolensed and therefore increase the number of expected femtolensed
events.

Although the source size requirements exclude the possibility of ever
detecting femtolensing in main sequence stars in nearby galaxies,
in the future, it may be possible to detect femtolensing of
white dwarfs.
(We are grateful to B. Paczy\'nski for pointing this out.)
In order for the fringes to have separation of 0.2 eV
(a typical optical bandwidth), the lensing masses would be in the range
$M \sim 10^{-9} - 10^{-11} M_{\sun}$
for short to characteristic time delays, respectively.
In the LMC, the maximum magnification possible for such events as determined
by the ratio of the Einstein ring to source size would be $\sim 2-20$ where
the larger masses cause the higher magnification.
For a white-dwarf source in the LMC observed at 1 eV,
the requirement that the source be smaller than the Fresnel length,
$(\lambda D)^{1/2} = 3\times 10^{9}$ is met ($D=D_{LMC}/2 =25 \mbox{kpc}$).
These events would be quite
short ($\le 1$~minute), due to relative velocities and a small Einstein ring,
relative to current microlensing events (e.g. \cite{p:86}).
However, it is not yet possible
to monitor white dwarves in the LMC, so one would have to
adopt an observing strategy for transients---possibly in the
ultraviolet---similar to that of supernovae searches. In contrast to
current microlensing studies, detection techniques
would be based on anti-correlation of the flux variation in
different wavelength bands.

\acknowledgements

This work was supported in part by the David \& Lucille Packard Foundation
and NASA grant NAG5-1901.
It is a pleasure to acknowledge B. \paczy ~for suggesting the white dwarf
lensing scenario as well as E.E. Fenimore, K.Z. Stanek, and H. Witt
for discussions.

\newpage

\section{Appendix A: Normalizations}

In this appendix, we describe the normalized units used in the
formulae above. We begin with the time-delay formula for a point mass
with shear (e.g. \cite{bn:92}):
\begin{equation}
\label{timedel_un}
\tilde{\tau}(\tilde{x},\tilde{y},\tilde{\mu},x_{\rm s},y_{\rm s})=
\frac{1+z_l}{c} \left[
({1\over 2D}+\tilde{\mu})\tilde{x}^2 + ({1\over 2D}-\tilde{\mu})\tilde{y}^2 -
{[x_{\rm s} \tilde{x} + y_{\rm s} \tilde{y}]\over d_{ls}}
-R_{\rm Sch} \ln(\tilde{x}^2+\tilde{y}^2) \right],
\end{equation}
where $z_l$ is the lens redshift,
$\tilde{x}, \tilde{y}$ are points at which the ray intersects the lens plane,
$x_{\rm s}, y_{\rm s}$ give the location of the source in the source plane,
$R_{\rm Sch}$ is the Schwarzschild radius, and
$D={d_ld_{ls}/d_s}$ where $d_{\rm l}, d_{\rm s}, d_{\rm ls}$
are angular diameter distances.

Throughout the derivation, additive constants will be ignored.
The form for $\Psi(\omega)$ is
\begin{equation}
\label{psiagain}
\Psi(\tilde{\omega})= \tilde{C}_{\tilde{\omega}} \int^{\infty}_{-\infty}
\int^{\infty}_{-\infty} {\rm d}\tilde{x} {\rm d}\tilde{y}
\exp(i\tilde{\omega}\tilde{\tau}).
\end{equation}
Then, we define the following dimensionless variables:
\begin{equation}
\label{tautilde}
\mbox{time delay:}~~~\tau \equiv {\tilde{\tau}c \over (1+z_l)R_{\rm Sch}}
\end{equation}
\begin{equation}
\label{omega}
\mbox{frequency:}~~~\omega \equiv {\tilde{\omega}(1+z_L)R_{\rm Sch}
\over c} = \frac{2\pi(1+z_L)R_{\rm Sch}}{\lambda}
\end{equation}
\begin{equation}
\label{x1y1}
\mbox{position (image plane):}~~~(x,y) \equiv {(\tilde{x},\tilde{y})
\over \sqrt{2DR_{\rm Sch}}}
\end{equation}
\begin{equation}
\label{mu}
\mbox{shear:}~~~\mu \equiv 2DR_{\rm Sch}\tilde{\mu};~~~0 \le \mu \le 1.
\end{equation}
With these definitions Eq. \ref{psiagain} becomes Eq. \ref{psi}, and
$C_{\omega}$ which normalizes the magnification to a unit flux,
can then be shown to be,
\begin{equation}
\label{C_w}
C_{\omega} = {i\omega \over \pi}
\end{equation}
Equation \ref{tautilde} can be rewritten after substituting the dimensionless
variables as
\begin{equation}
\label{halftau}
\tau(x,y,\mu,x_s,y_s) = (1 + \mu)x^2 + (1 - \mu)y^2 -
{\sqrt{2D}(x_sx + y_sy) \over d_{ls}\sqrt{R_{\rm Sch}}} - \ln(x^2+y^2).
\end{equation}
Finally, we choose an impact parameter
\begin{equation}
\label{theta}
\theta \equiv {\sqrt{x_s^2+y_s^2} \over d_{\rm s}R_{\rm E}},
\end{equation}
where $R_{\rm E}$ is the Einstein radius:
\begin{equation}
\label{einstein}
R_{\rm E} \equiv \sqrt{2R_{\rm Sch}d_{\rm ls} \over d_{\rm l}d_{\rm s}}.
\end{equation}
We also pick $\phi$ so that we recover Eq. \ref{timedel}
\begin{equation}
\label{timedel1}
\tau(x,y,\mu,\phi,\theta)=(1+\mu)x^2 + (1-\mu)y^2 - 2\theta[x\cos(\phi) +
y\sin(\phi)] - \ln(x^2+y^2).
\end{equation}

\section{Appendix B: Semi-Classical Approximation}

Herein, we discuss the behavior of
$\tilde{\Psi}(t)$ near critical points (images), so that the
divergent portions of the function can be subtracted to yield a
smooth function that can be computationally sampled and transformed.
When the frequency is larger than the minimum separation of image
time delays, ie.
$\omega \gg |\tau_{\rm i}-\tau_{\rm j}|$ for all i, j,
it is in the semi-classical region and these singularities
dominate $\Psi(\omega)$.
In any regime, the singular behavior dominates
$\tilde{\Psi}(t)$ near the time delays of the images.

We consider the two cases of a minimum/maximum and a saddle point:

\subsection{Minimum/Maximum}

Near a critical point at the origin, a contour is given by
\begin{equation}
\label{tau_cont}
\tau = \pm({x^2 \over 2a^2}+{y^2 \over 2b^2}) +t_{\rm crit},
\end{equation}
where $t_{\rm crit}$ is the time delay at the critical point $x = y = 0$.
Without loss of generality we consider the case of a minimum only.
The contribution of the singularity to $\tilde{\Psi}(t)$
is found by the integral (see eqs. \ref{minmax}--\ref{sumlines})
\begin{equation}
| \oint du |~~~\mbox{where}~~~{\pard x\over \pard u} =
-{\pard \tau\over
\pard y},~~~{\pard y\over \pard u} = {\pard \tau\over \pard x}.
\end{equation}
We write:
\begin{equation}
{\pard^2 x\over \pard u^2} = -({\pard \over \pard u})({\pard \tau \over
\pard y}) =  -({\pard \over \pard u})({y\over b^2}) =
-{x\over a^2b^2},~~~\mbox{so that}
\end{equation}
\begin{equation}
x=A\cos({u \over ab} + \phi)~~~\mbox{and}
\end{equation}
\begin{equation}
\oint du = 2\pi a b.
\end{equation}
Further, from eq.  \ref{tau_cont}
\begin{equation}
{1\over a^2b^2} = {\pard^2\tau\over \pard x^2}{\pard^2\tau\over \pard y^2} -
({\pard^2\tau\over \pard x\pard y})^2 \equiv \det H.
\end{equation}
The singularity is such therefore, that
\begin{equation}
\Delta \tilde{\Psi}(t) = \left\{ \begin{array}{ll}
	0 & \mbox{if}~~~t < t_{\rm min} \\
	2\pi/\sqrt{\det H} & \mbox{if}~~~t > t_{\rm min}
\end{array}
\right.
\end{equation}
This term can be removed from the slowly varying part of $\tilde{\Psi(t)}$
and Fourier transformed to yield the semi-classical contribution as
\begin{equation}
\Delta{\Psi}(\omega)= C_{\omega}\int dt \exp(i\omega
t)\Delta\tilde{\Psi}  =
{-2\exp(i\omega t_{\rm min})\over \sqrt{\det H}}
\end{equation}

\subsection{Saddle Point}
In a similar manner as above, the singularity of a saddle point may be
calculated. Without loss of generality we consider this orientation of
saddle point:
\begin{equation}
\label{tau_cont1}
\tau = ({x^2 \over 2a^2}-{y^2 \over 2a^2}).
\end{equation}
By analogy with the last derivation, we have
\begin{equation}
x=A\cosh(u/ab)
\end{equation}
In this case the contours are hyperbolic rather than
elliptical and do not close locally.
To compute the contribution from
the neighborhood of the critical point,
we consider that part of each contour for which $|x| < x_o$, where
$x_o$ is fixed as $\tau\to t_{\rm crit}$.
Finding $u_o$ such that $x(u_o) = x_o$, and taking
${|u_o / ab| \gg 1}$, we have
\begin{equation}
u_o = ab\cosh^{-1}({u_o \over ab}) \to {1 \over 2} a b \log({2 x_o \over A}) =
-{1\over 2} a b
\log|t-t_{\rm crit}| + \mbox{constant},
\end{equation}
where the constant depends on the choice of $x_o$.
The contour extends from $-u_o$ to $u_o$ on each of 2 branches of the
hyperbola, so
\begin{equation}
\Delta \tilde{\Psi}(t) = -2ab \log |t - t_{\rm crit}| =
{-2\log | t - t_{\rm crit} | \over \sqrt{-\det H}}.
\end{equation}
When Fourier transformed, $\tilde{\Psi(t)}$
yields the semi-classical contribution,
\begin{equation}
\Delta{\Psi}(\omega)= C_{\omega}\int dt \exp(i\omega t)\Delta\tilde{\Psi} =
{2i\exp(i\omega t_{\rm crit})\over \sqrt{-\det H} }
\end{equation}

\section{Appendix C: Physical Optics of Merging Images}

When the flux is dominated by two bright images near a critical line,
the time-delay surface can be approximated locally by
\begin{equation}
\label{tau_fold}
\tau(x,y;x_s,y_s)= \frac{x^3}{3a^3} +\frac{y^2}{b^2} -x_s x -y_s y.
\end{equation}
Here $(x_s,y_s)$ is the undeflected source position in the lens plane.
We use the dimensionless units of Appendix A, but the local coordinates
in equation (\ref{tau_fold}) are centered on the critical line ($x=0$)
rather than the deflecting mass.
The parameters $a$ and $b$ are assumed to be $\sim 1$, but their precise
values depend on the global geometry of the time-delay surface.

For $x_s>0$, geometric optics yields two images at
$x=\pm a^{3/2}x_s^{1/2}$ with net flux
\begin{equation}
\label{F_geo}
F_{\rm geo}(x_s) = 2a^{3/2}b^2 x_s^{-1/2}.
\end{equation}
On the other hand, the amplitude (\ref{psi}) can be evaluated directly:
\begin{equation}
\label{F_phys}
F(x_s,\omega)= |\psi(\omega;x_s,y_s)|^2= 4\pi\omega^{1/3}a^2b^2
|\Ai(-ax_s\omega^{2/3})|^2.
\end{equation}
The Airy function quickly approaches the
asymptotic approximation [e.g., \cite{AS:72}]
\begin{equation}
\label{Airy}
\Ai(-\xi)\approx\pi^{-1/2}\xi^{-1/4}
\sin\left(\frac{2}{3}\xi^{3/2}+\frac{\pi}{4}\right)~~~\mbox{if $\xi\ge 1$},
\end{equation}
so that the flux (\ref{F_phys}) displays fringes with spacing
\begin{equation}
\label{fringes}
\Delta\omega_{\mbox{f}} = \frac{3\pi}{2}(ax_s)^{-3/2};
\end{equation}
in fact, even the position of first zero in the flux is accurately
predicted by the asymptotic form (\ref{Airy}), which is the
semiclassical prediction.
Both geometric optics and the semiclassical theory predict a divergent
flux as $x_s\to 0$.
But since $\Ai(0)= 3^{-2/3}/\Gamma(2/3)$,
the flux at the caustic is finite and scales as $\omega^{1/3}$.

Comparing (\ref{F_geo}) and (\ref{fringes}), we see that {\it the
fringe spacing scales as the cube of the total magnification}.
This accounts for the long-period fringes that we often see in our
numerical spectra.
Because of magnification bias, observed fringes should typically
be much more widely spaced than the naive scaling
$\Delta\tilde\omega\sim c/R_{\rm Sch}$ assumed in earlier work.

A source of finite angular size can be regarded
as an ensemble of incoherent points.
If fringes are to be visible, the phase of the Airy function in
(\ref{F_phys}) must vary by less than the fringe spacing (\ref{fringes})
across the source.
Therefore the angular size must satisfy
\begin{equation}
\label{source_size}
\theta_s \lesssim \theta_E\frac{0.6}{a}\left(
\frac{\Delta\omega_{\mbox{f}}}{\omega}\right)\omega^{-2/3},
\end{equation}
where $\theta_E\equiv \sqrt{2DR_{\rm Sch}/d_l}$ is the angular
radius of the Einstein ring.
For images close to a critical line, this limit supersedes
the angular-size limit associated with the
Fresnel length, which scales as $\omega^{-1/2}$.
To achieve fringes of a given physical frequency $\tilde\omega$
and spacing $\Delta\tilde\omega$ with lenses whose Schwarzschild
radius is much larger than the wavelength, the limit on the source size
scales as $R_{\rm Sch}^{-1/6}$.

\newpage

\end{document}